# Validation and Improvement of Data Assimilation for Flood Hydrodynamic Modelling Using SAR Imagery Data

Thanh Huy Nguyen[1], Anthéa Delmotte[1], Christophe Fatras[2], Peter Kettig[3], Andrea Piacentini[1], and Sophie Ricci[1]

[1] CECI, CERFACS/CNRS UMR 5318, Toulouse, France
[2] Collecte Localisation Satellites (CLS), Toulouse, France
[3] Centre National d'Etudes Spatiales (CNES), Toulouse, France

Corresponding author: thnguyen@cerfacs.fr

*Abstract*—Relevant comprehension of flood hazards has emerged as a crucial necessity, especially as the severity and the occurrence of flood events may intensify with climate changes. Flood simulation and forecast capability have been greatly improved thanks to advances in data assimilation. This approach combines in-situ gauge measurements with hydrodynamic models, which aims at correcting the hydraulic states and reducing the uncertainties in the model parameters, e.g., friction coefficients, inflow discharge. These methods depend strongly on the availability and quality of observations, thus requiring other data sources to improve the flood simulation and forecast quality. Sentinel-1 images collected during a flood event were used to classify an observed scene into dry and wet areas. The study area concerns the Garonne Marmandaise catchment, and focuses on the recent flood event in January-February 2021. In this paper, seven experiments are carried out, two in free run modes (FR1 and FR2) and five in data assimilation modes (DA1 to DA5). A model-observation bias was diagnosed and corrected over the beginning of the flood event. Quantitative assessments are carried out involving 1D metrics at Vigicrue observing stations and 2D metrics with respect to the Sentinel-1 derived flood extent maps. They demonstrate improvements on flood extent representation thanks to the data assimilation and bias correction.

## I. Introduction

Simulations based on hydrodynamic numerical models in analysis and forecast modes are crucial to mitigate flood impacts. The analysis mode could be carried out to obtain better estimates of the dynamic footprints of past flood events, as well as to assess flood damages and design future flood defense systems, whereas the forecast mode is used by civil security services and industry. However, these numerical models remain imperfect because the uncertainties inherently existing within the models and the inputs, e.g., friction and boundary conditions, seemingly translate into uncertainties in the model outputs. A well-established method for reducing uncertainties and generating more reliable predictions is to periodically adjust these models, for instance, by assimilating various observations as they become available [1].

Indeed, flood simulation and forecast capability have been greatly improved thanks to the advances in data assimilation (DA). Such methods, notably Ensemble Kalman Filter (EnKF), aim at combining in-situ gauge measurements with numerical models to correct the hydraulic states and reduce the uncertainties in the model parameters (e.g., friction coefficients, upstream inflow). These filters rely on the stochastic computation of the forecast error covariance matrix, within a limited number of simulations. The sources of uncertainty, represented by the control vector, is updated over each assimilation window. Nevertheless, this approach depends strongly on the availability and quality of observations, as its performance relies on the spatial and temporal density of the observing network [2]. As a matter of fact, limnimetric in-situ observations providing water levels are only available at a few sparse locations along a river catchment, due to installation and maintenance costs [3]. This is a limiting factor for numerical model precision in simulation and forecast, especially in the floodplains. Such a situation requires efforts to leverage other sources of data such as remote sensing-derived flood maps to validate and improve the flood simulation and forecasting performance. In this work, we carry out the flood extent mapping by applying a Random Forest (RF) segmentation on Synthetic Aperture Radar (SAR) images such as Sentinel-1 (S1) [4]. The inferred flood extent maps are then compared with the flood extents simulated by TELEMAC-2D with EnKF assimilation.

This work highlights the merits of using SAR-derived flood extent maps to validate and improve the simulation results based on hydrodynamic numerical models with EnKF DA. It illustrates how SAR imagery data could be used to overcome the limits of the calibration and validation process which was done using river-gauge data only. For instance, a bias between the models and in-situ observations has been identified and corrected, yielding better flood extent representation. Quantitative performance assessments are carried out by comparing the simulated and observed water level time-series at several in-situ gauge locations, as well as involving Critical Success Index measured between the simulated flood extent maps and the SAR-derived maps. They underline the benefits of using spatially distributed remote sensing data that inform on the floodplain dynamics.

## II. Study Area, Data, Model

Hydrodynamic numerical models, such as TELEMAC-2D (www.opentelemac.org), are used to simulate and predict water surface elevation and velocity from which the flood risk can be assessed for lead times ranging from a couple of hours to several days. TELEMAC-2D solves the Shallow Water Equations (SWE) with an explicit first-order time integration scheme, a finite element scheme and an iterative conjugate gradient method [5]. At each point within the mesh representing the model topography and bathymetry (for mesh nodes in the river channel), the results of the simulation are water depth and velocity averaged over the azimuth axis.

### A. Shallow Water Equations (SWEs) in TELEMAC-2D

The non-conservative form of SWEs is written in terms of water depth ($h$ [m], also called water level) and horizontal components of velocity ($u$ and $v$ [m.s$^{-1}$]). They express mass and momentum conservation averaged in the vertical dimension while assuming that:
- The horizontal length scale is significantly greater than the vertical scale;
- Vertical pressure gradients are hydrostatic;
- Horizontal pressure gradients are due to the displacement of the free surface.

As such, using similar mathematic notations to [6], the SWEs read:

$$\frac{\partial h}{\partial t} + \frac{\partial}{\partial x}(hu) + \frac{\partial}{\partial y}(hv) = 0 \quad (1)$$

$$\frac{\partial u}{\partial t} + u\frac{\partial u}{\partial x} + v\frac{\partial u}{\partial y} = -g\frac{\partial z}{\partial x} + F_x + \frac{1}{h}div\left(hv_e\overrightarrow{grad}(u)\right) \quad (2)$$

$$\frac{\partial v}{\partial t} + u\frac{\partial v}{\partial x} + v\frac{\partial v}{\partial y} = -g\frac{\partial z}{\partial y} + F_y + \frac{1}{h}div\left(hv_e\overrightarrow{grad}(v)\right) \quad (3)$$

where $z$ [m NGF69] is the water surface elevation ($h = z - z_b$ with $z_b$ [m NGF69] being the bottom elevation) and $v_e$ [m$^2$.s$^{-1}$] is the water diffusion coefficient. $g$ [m.s$^{-2}$] is the gravitational acceleration constant. $div$ and $\overrightarrow{grad}$ are respectively the divergence and gradient operators. $F_x$ and $F_y$ [m.s$^{-2}$] are the horizontal components of external forces (friction, wind and atmospheric forces), defined as follows:

$$\begin{cases} F_x = -\frac{g}{K_s^2}\frac{u\sqrt{u^2+v^2}}{h^{4/3}} - \frac{1}{\rho_w}\frac{\partial P_{atm}}{\partial x} \\ \quad\quad\quad + \frac{1}{h}\frac{\rho_{air}}{\rho_w}C_D U_{w,x}\sqrt{U_{w,x}^2 + U_{w,y}^2} \\ F_y = -\frac{g}{K_s^2}\frac{v\sqrt{u^2+v^2}}{h^{4/3}} - \frac{1}{\rho_w}\frac{\partial P_{atm}}{\partial y} \\ \quad\quad\quad + \frac{1}{h}\frac{\rho_{air}}{\rho_w}C_D U_{w,y}\sqrt{U_{w,x}^2 + U_{w,y}^2} \end{cases}$$

where $\rho_w/\rho_{air}$ [kg.m$^{-3}$] is the water/air density, $P_{atm}$ [Pa] is the atmospheric pressure, $U_{w,x}$ and $U_{w,y}$ [m.s$^{-1}$] are the horizontal wind velocity components, $C_D$ [-] is the wind drag coefficient that relates the free surface wind to the shear stress, and lastly, $K_s$ [m$^{1/3}$.s$^{-1}$] is the river bed and floodplain friction coefficient using the Strickler formulation [7]. In order to solve Eq. (1)-(3), initial conditions $\{H(x,y,t=0) = H_0(x,y); u(x,y,t=0) = u_0(x,y); v(x,y,t=0) = v_0(x,y)\}$ are provided, and boundary conditions (BC) are described with a time-dependent hydrograph upstream and a rating curve downstream. The Strickler coefficient is prescribed as uniform over subdomains, and calibrated according to the observing network.

### B. Study area

The study area concerns the Garonne Marmandaise catchment (Southwest France) which extends over a 50-km reach of the Garonne River, between Tonneins, downstream of the confluence with the river Lot, and La Réole. This part of the valley is identified as an area at high flood risk. Since the 19th century, it has been equipped with infrastructures to protect the Garonne floodplain from flooding events such as the historic flood of 1875. A system of longitudinal dykes and weirs was progressively constructed to protect floodplains and manage submersion and flood retention areas.

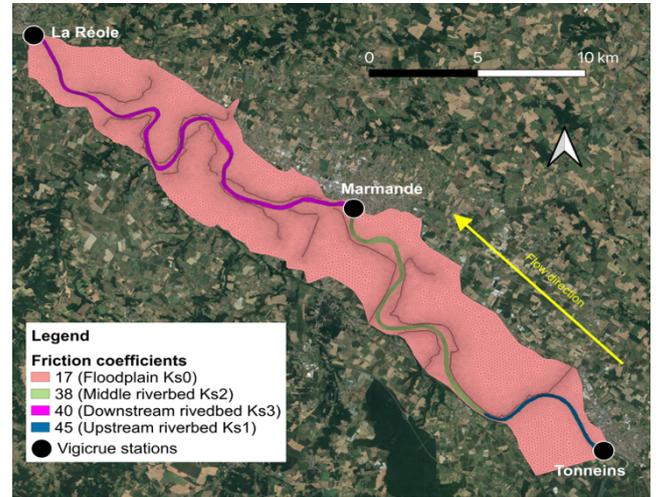
(a) Mesh and friction zoning

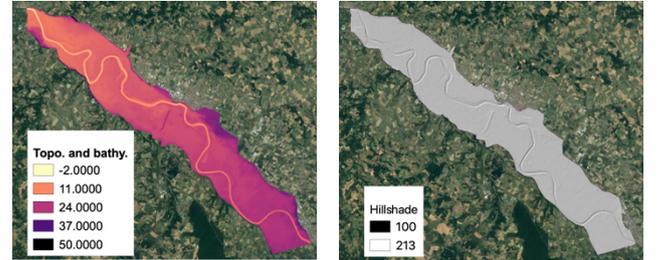
(b) Topography and bathymetry    (c) Hillshade representation
Figure 1: Garonne Marmandaise TELEMAC-2D model.

A TELEMAC-2D model (Figure 1) was developed and calibrated over this catchment, which was built on a mesh of 41,000 nodes using bathymetric cross-sectional profiles and topographic data [8]. It involves a triangular unstructured mesh, with an increased mesh resolution around the dykes and in the river bed. The local rating curve at Tonneins (established from a number of water level-discharge measurements) translate the observed water levels into a discharge that is then applied over the entire upstream interface, both river bed and floodplain boundary cells. This modeling strategy was implemented by Electricité de France R&D to allow a cold start of the model with any inflow value. However, it prompts an over-flooding of the upstream first meander (near Tonneins), until the water returns to the river bed. The downstream BC at La Réole is described with a local rating curve. Over the simulation domain, the friction coefficient $K_s$ is defined over four areas. Their values resulted from a calibration procedure over a set of non-overflowing events and are set respectively equal to: $K_{s_1} = 45$, $K_{s_2} = 38$ and $K_{s_3} = 40$ [m$^{1/3}$s$^{-1}$] for the upstream, middle and downstream part of the river bed and $K_{s_0} = 17$ [m$^{1/3}$s$^{-1}$] for the floodplain. They are characterized by a discrete zoning of uniform $K_s$ values into subdomains within the catchment, restricted by the limited number of in-situ measurements. Such a friction coefficient setting is indeed prone to uncertainty related to the zoning assumption, the calibration procedure and the set of calibration events. This uncertainty is more significant in the floodplain area where no observing station is available.

The probability density function (PDF) for the Strickler coefficients is assumed to follow a gaussian distribution with mean and standard deviation set accordingly to the calibration process and expert knowledge. The limited number of in-situ observations also yields errors in upstream inflow as the expression of the inflow relies on the use of the local rating curve, usually involves extrapolation for high flows. In order to account for uncertainties in the upstream BC (i.e., time-dependent discharge $Q_{up}(t)$) while limiting the dimension of the uncertain input space, the perturbation added to BC is applied via a parametric formulation that allows for a multiplicative, an additive and a time-shift error, as proposed by [9]:

$$\tilde{Q}_{up}(t) = a \times Q_{up}(t - c) + b \qquad (4)$$

where $(a, b, c) \in \mathbb{R}^3$, and their PDF follows gaussian distribution, centered at their default values. The characteristics of the friction- and inflow-related uncertainty PDFs are summarized in Table 1. Other works dealing with this uncertainty have been put forth using EnKF [10] or Extended Kalman filter [11].

Table 1: Gaussian PDF of uncertain input variables related to friction and inflow discharge coefficients.

| Variables | Unit | Calibrated/default value $\mathbf{x}_0$ | Standard deviation $\sigma_\mathbf{x}$ | 95% confidence interval |
|---|---|---|---|---|
| $K_{s_0}$ | $m^{1/3}s^{-1}$ | 17 | 0.85 | $17 \pm 1.67$ |
| $K_{s_1}$ | $m^{1/3}s^{-1}$ | 45 | 2.25 | $45 \pm 4.41$ |
| $K_{s_2}$ | $m^{1/3}s^{-1}$ | 38 | 1.9 | $38 \pm 3.72$ |
| $K_{s_3}$ | $m^{1/3}s^{-1}$ | 40 | 2.0 | $40 \pm 3.92$ |
| $a$ | - | 1 | 0.06 | $1 \pm 0.118$ |
| $b$ | $m^3s^{-1}$ | 0 | 100 | $0 \pm 196$ |
| $c$ | s | 0 | 900 | $0 \pm 1760$ |

### C. 2021 flood event and observations

A substantial flood event occurred in late January and February 2021 as it exceeded the yellow risk level alert set out by the French national flood forecasting center (SCHAPI) in collaboration with the departmental prefect, and reached its peak on February 4. In this work, we examine an extended length of this event, i.e., between January 16 and February 15.

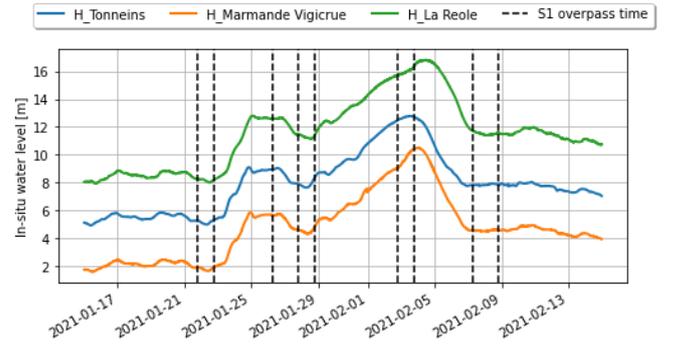

Figure 2: Water level time series at Vigicrue observing stations, and Sentinel-1 overpass times.

Figure 2 depicts the in-situ water level (15-minute time step) observed during the flood event at Vigicrue observing stations: Tonneins (blue curve), Marmande (orange curve) and La Réole (green curve). The event is observed by nine S1 images, indicated by the vertical black dashed lines in Figure 2. S1 works as a constellation of two satellites in a phased orbit, S1A and S1B, each with a 6-day revisit frequency. They are part of the Copernicus program launched by ESA with contributions from CNES. The flood peak was covered by the ascending orbit 30 on February 2 18:55 and by the ascending orbit 132 the next day on February 3 18:48.

### III. METHODS

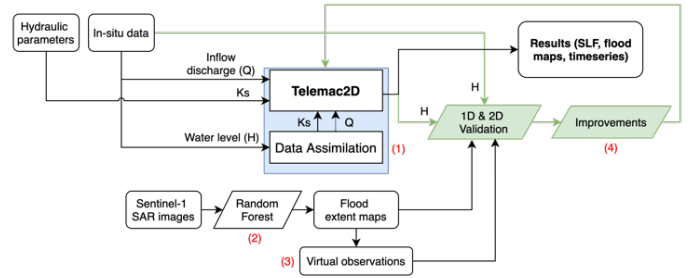

Figure 3: Proposed methodology.

The merits of the complementary use of in-situ data for assimilation and remote sensing (RS) data for validation are assessed with different deterministic and ensemblist numerical experiments. Four main steps to sum up the dynamic and novelty of the proposed methodology (Figure 3):

(1) DA combines in-situ data with this TELEMAC2D model to reduce the uncertainties within the model and inputs. The control variables are the friction coefficients and inflow parametric coefficients;

(2) S1 imagery data is used to generate flood extent maps, represented by binary flood maps (black pixels: dry, white pixels: wet);

(3) Several virtual observation regions are defined with expertise at important areas within the simulation domain. They measure the number of wet pixels within defined boxes and allow to validate local performance of the flood simulations;

(4) Based on the performed 1D and 2D validations, improvements are made upon the model and simulation setting, allowing to better configure the control space, correct the existing bias between the model and observation, and so on.

*A. Ensemble-based data assimilation algorithm (EnKF)*

Continuous time-series of measured water levels and/or discharges recorded at discrete locations are traditionally used for model calibration and validation of DA algorithms for real-time constraints of hydraulic flood prediction models [1] [12]. In this work, the measured water levels at the 3 Vigicrue stations (Tonneins, Marmande and La Réole) are assimilated with the EnKF algorithm in the TELEMAC-2D Garonne model presented previously to sequentially correct the friction and inflow discharge.

*1) Description of the control vector:*

The DA algorithm consists in a cycled stochastic EnKF, where the control vector $\mathbf{x}$ is composed of the friction coefficients (four scalars $K_{s_i}, i \in 0,\ldots,3$) and parameters that modify the time-dependent upstream BC (three scalars $a, b, c$). $n$ denotes the size of the control vector. These seven parameters are assumed to be constant over a DA cycle, yet their evolution in time is made possible by DA between cycles. The DA cycle $k$ covers a time window, noted $W_k = [t_{start}, t_{end}]$ of 12-hour length over which $N_{obs}$ in-situ observations are assimilated. The cycling of the DA algorithms consists in sliding the window by a period $T_{shift} = $ 6 hours so that the cycles $W_k$ and $W_{k+1}$ overlap. The EnKF algorithm relies on the propagation of $N_e$ members with perturbed values of $\mathbf{x}$ (denoted by $\mathbf{x}^i$) into the forecast values denoted by $\mathbf{x}_k^{f,i}$ with $i \in [1, N_e]$ represents the ensemble member index.

*2) EnKF forecast step:*

The EnKF forecast step stands in the propagation in time of the control and model state vectors over the assimilation window $W_k$ that gathers $N_{obs}$ observations. The EnKF is here applied to model parameters that, by definition, do not evolve in time over the $W_k$. In order to avoid ensemble collapse, artificial dispersion is introduced within the sampling with the addition of perturbations $\boldsymbol{\theta}$ to the difference between the mean of the analysis from the previous cycle $(k-1)$ and the previous cycle analysis. The two terms are weighted by the hyperparameter $\lambda$. The forecast step thus reads:

$$\mathbf{x}_k^{f,i} = \begin{cases} \mathbf{x}_0 + \boldsymbol{\theta}_k^i & if\ k = 1 \\ \overline{\mathbf{x}_{k-1}^a} + \lambda_1\left(\mathbf{x}_{k-1}^{a,i} - \overline{\mathbf{x}_{k-1}^a}\right) + (1-\lambda_1)\boldsymbol{\theta}_k^i & if\ k > 1 \end{cases} \quad (5)$$

with $\overline{\mathbf{x}_{k-1}^a} = \left(\sum_{i=1}^{N_e} \mathbf{x}_{k-1}^{a,i}\right)/N_e \in \mathbb{R}^n$ and $\boldsymbol{\theta}_k^i \sim \mathcal{N}(\mathbf{0}, \sigma_\mathbf{x}^2)$.

For the first cycle, the perturbed friction and upstream forcing coefficient values are drawn within the PDFs described in Table 1. For the next cycles, the set of coefficients issued from the analysis at the previous cycle is further dispersed by combining the analysis anomalies with perturbations $\boldsymbol{\theta}$ drawn from the gaussian distribution centered at 0 and with the standard deviation described in Table 1. This technique is an alternative to anomalies inflation for avoiding the ensemble collapse, while preserving part of the information from the background statistical characteristics. In the following implementation, $\lambda$ is respectively set equal to 0.3. The background hydraulic state, denoted by $\mathbf{s}_k^{f,i}$, associated with each member of the ensemble results from the integration of the hydrodynamic model $\mathcal{M}_k: \mathbb{R}^n \to \mathbb{R}^m$ from the control space to the model state (of dimension $m$) over $W_k$:

$$\mathbf{s}_k^{f,i} = \mathcal{M}_k\left(\mathbf{s}_{k-1}^{a,i}, \mathbf{x}_k^{f,i}\right) \quad (6)$$

The initial condition for $\mathcal{M}_k$ at $t_{start}$ is provided by a user-defined restart file for the first cycle. For the following cycles, it stems from the analyzed model state $\mathbf{s}_{k-1}^{a,i}$, saved from the previous cycle. The control vector equivalent in the observation space for each member, noted $\mathbf{y}_k^{f,i}$, stems from:

$$\mathbf{y}_k^{f,i} = \mathcal{H}_k\left(\mathbf{s}_k^{f,i}\right) \quad (7)$$

where $\mathcal{H}_k: \mathbb{R}^m \to \mathbb{R}^{N_{obs}}$ is the observation operator from the model state space to the observation space (of dimension $N_{obs}$) that selects, extracts and eventually interpolates model outputs at times and locations of the observation vector $\mathbf{y}_k^o$ over $W_k$. It should be noted that, in the following, the observation operator may also include a bias correction to take into account a systematic model error. Eq. (7) thus reads

$$\mathbf{y}_k^{f,i} = \mathcal{H}_k\left(\mathbf{s}_k^{f,i}\right) - \mathbf{y}_{bias} \quad (8)$$

where $\mathbf{y}_{bias}$ is an a priori knowledge of the model-observation bias.

*3) EnKF analysis step:*

The EnKF analysis step stands in the update of the control and model state vectors. When applying a stochastic EnKF [13], the observation vector $\mathbf{y}^{o,i}$ is perturbed, thus an ensemble of observations $\mathbf{y}_k^{o,i}$ ($i \in [1, N_e]$) is generated:

$$\mathbf{y}_k^{o,i} = \mathbf{y}_k^o + \boldsymbol{\epsilon}_k \text{ with } \boldsymbol{\epsilon}_k \sim \mathcal{N}(0, \mathbf{R}_k) \quad (9)$$

where $\mathbf{R}_k = \sigma_{obs}^2 \mathbf{I}_{N_{obs}}$ is the observation error covariance matrix, here assumed to be diagonal, of standard deviation $\sigma_{obs}$ (and $\mathbf{I}_{N_{obs}}$ is the $N_{obs} \times N_{obs}$ identity matrix), as the observation errors are assumed to be uncorrelated, Gaussian and with a standard deviation proportional to the observations $\sigma_{obs,k} = \tau y_k^o$. The innovation vector over $W_k$ is the difference between the perturbed observation vector $\mathbf{y}_k^{o,i}$ and the model equivalent $\mathbf{y}_k^{f,i}$ from Eq. (7) (or Eq. (8)) and Eq. (9). It is weighted by the Kalman gain matrix $\mathbf{K}_k$ and then added as a correction to the background control vector $\mathbf{x}_k^{f,i}$, so that the analysis control vector $\mathbf{x}_k^{a,i}$ is computed in Eq. (10),

$$\mathbf{x}_k^{a,i} = \mathbf{x}_k^{f,i} + \mathbf{K}_k\left(\mathbf{y}_k^{o,i} - \mathbf{y}_k^{f,i}\right) \quad (10)$$

The Kalman gain reads:

$$\mathbf{K}_k = \mathbf{P}_k^{x,y}\left[\mathbf{P}_k^{y,y} + \mathbf{R}_k\right]^{-1} \quad (11)$$

with $\mathbf{P}_k^{y,y}$ being the covariance matrix of the error in the background state equivalent in the observation space $\mathbf{y}_k^f$ and $\mathbf{P}_k^{x,y}$ the covariance matrix between the error in the control vector and the error in $\mathbf{y}_k^f$, stochastically estimated within the ensemble:

$$\mathbf{P}_k^{x,y} = \frac{1}{N_e}\mathbf{X}_k^T \mathbf{Y}_k \in \mathbb{R}^{n \times N_{obs}}$$

$$\mathbf{P}_k^{y,y} = \frac{1}{N_e}\mathbf{Y}_k^T \mathbf{Y}_k \in \mathbb{R}^{N_{obs} \times N_{obs}}$$

with

$$\mathbf{X}_k = \left[\mathbf{x}_k^{f,1} - \overline{\mathbf{x}_k^f}, \cdots, \mathbf{x}_k^{f,N_e} - \overline{\mathbf{x}_k^f}\right] \in \mathbb{R}^{n \times N_e}$$

$$\mathbf{Y}_k = \left[\mathbf{y}_k^{f,1} - \overline{\mathbf{y}_k^f}, \cdots, \mathbf{y}_k^{f,N_e} - \overline{\mathbf{y}_k^f}\right] \in \mathbb{R}^{N_{obs} \times N_e}$$

where

$$\overline{\mathbf{x}_k^f} = \frac{1}{N_e}\sum_{i=1}^{N_e}\mathbf{x}_k^{f,i} \in \mathbb{R}^n \text{ and } \overline{\mathbf{y}_k^f} = \frac{1}{N_e}\sum_{i=1}^{N_e}\mathbf{y}_k^{f,i} \in \mathbb{R}^{N_{obs}}$$

The analyzed hydrodynamic state, associated with each analyzed control vector $\mathbf{x}_k^{a,i}$ is denoted by $\mathbf{s}_k^{a,i}$. It results from the integration of the hydrodynamic model $\mathcal{M}_k$ with updated friction and upstream forcing over $W_k$, starting from the same initial condition (for the first cycle), then from each background simulation within the ensemble:

$$\mathbf{s}_k^{a,i} = \mathcal{M}_k\left(\mathbf{s}_{k-1}^{a,i}, \mathbf{x}_k^{a,i}\right) \quad (12)$$

*B. Flood extent mapping using S1 images*

In recent years, SAR image data has been widely used in flood management due to its ability to collect day and night images in all weather and to map flood extents in large areas in near real-time. Water bodies and flooded areas usually appear on SAR images with low backscatter intensity because most of the incident radar signals are reflected away from the SAR antenna. Therefore, with a few exceptions such as the built environment and vegetation areas, the detection of such areas is straightforward on SAR images.

In this work, a Random Forest (RF) segmentation [14] was trained over a training dataset with 223 S1 images from 12 non-coastal Copernicus EMS Rapid mapping flood cases from multiple regions of the world. The training dataset consists of permanent water pixel samples selected according to the Global Surface Water Occurrence products [15].

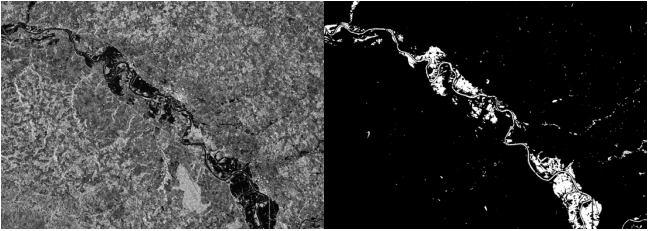

Figure 4: SAR S1 image (left) acquired on 2021-02-02 and the flood extent map inferred by RF (right).

Both VV and VH polarizations of S1 were used, in conjunction with the use of the local slope derived from the MERIT DEM [16]. The S1 images are calibrated and orthorectified, then inferred by the RF algorithm to produce the binary flood map, indicating flood (white pixels) and non-flood areas (black pixels), as shown in Figure 4. Using cuML, an open-source GPU-accelerated machine learning library, the RF algorithm is able to generate a flood extent map in a couple of minutes. To remove noises and artifacts in the resulting detected binary flood maps, a majority filter (with size of 3) was applied on the resulting flood binary map. The ground sampling distance of the S1 images and the derived flood binary map is 10 x 10 meters. The generated flood extent maps are used to improve flood visualization and reduce modeling uncertainty.

*C. Experimental setup*

Two free run simulations and five DA simulations were implemented, summarized by Table 2. Their variations concerns whether the experiment consists in a Free Run or involves a DA approach, as well as whether or not the model-observation bias $\mathbf{y}_{bias}$ is taken into account, and the value of $\tau$ (for $\sigma_{obs}$). The diagnosed bias was estimated during the 24 hours of January 15 which are composed of quasi-stationary non-overflowing discharge, result in $\mathbf{y}_{bias,Tonneins} = 0.72$, $\mathbf{y}_{bias,Marmande} = 0.4$, and $\mathbf{y}_{bias,LR} = -0.24$ meters.

Table 2: Summary of the realized experiments.

| Exp. name | Bias correction | DA | $N_e$ | $\tau$ (%) | Control variables |
|---|---|---|---|---|---|
| FR1 | No | No | 1 | - | - |
| FR2 | Yes | No | 1 | - | - |
| DA1 | No | Yes | 24 | 15 | $K_{s[0:3]}, a, b, c$ |
| DA2 | Yes | Yes | 24 | 15 | $K_{s[0:3]}, a, b, c$ |
| DA3 | Yes | Yes | 24 | 1 | $K_{s[0:3]}, a, b, c$ |
| DA4 | Yes | Yes | 24 | 99 | $K_{s[0:3]}, a, b, c$ |
| DA5 | Yes | Yes | 24 | 15 | $K_{s[0:3]}$ |

IV. RESULTS AND DISCUSSIONS

In terms of validation, two assessments are carried out. First, a 1D validation between the simulated and observed time-series water levels at Vigicrue observing stations was achieved. Second, 2D assessments between the flood extent maps simulated by TELEMAC-2D and those derived from Sentinel-1 images by RF algorithm was carried out in order to validate the overall performance of simulated flood extents, and to analyze the local behavior at individual virtual observation locations.

*A. Simulated water levels*

Figure 5 depicts the water levels simulated by the performed experiments (from FR1 to DA5) represented by solid curve with respect to the observed water levels (black dashed curve) at the 3 Vigicrue observing stations. These plots focus on the period near the flood peak, i.e., between January 31 and February 9. The bottom panel on each sub-figure shows the differences between the simulated and observed water levels.

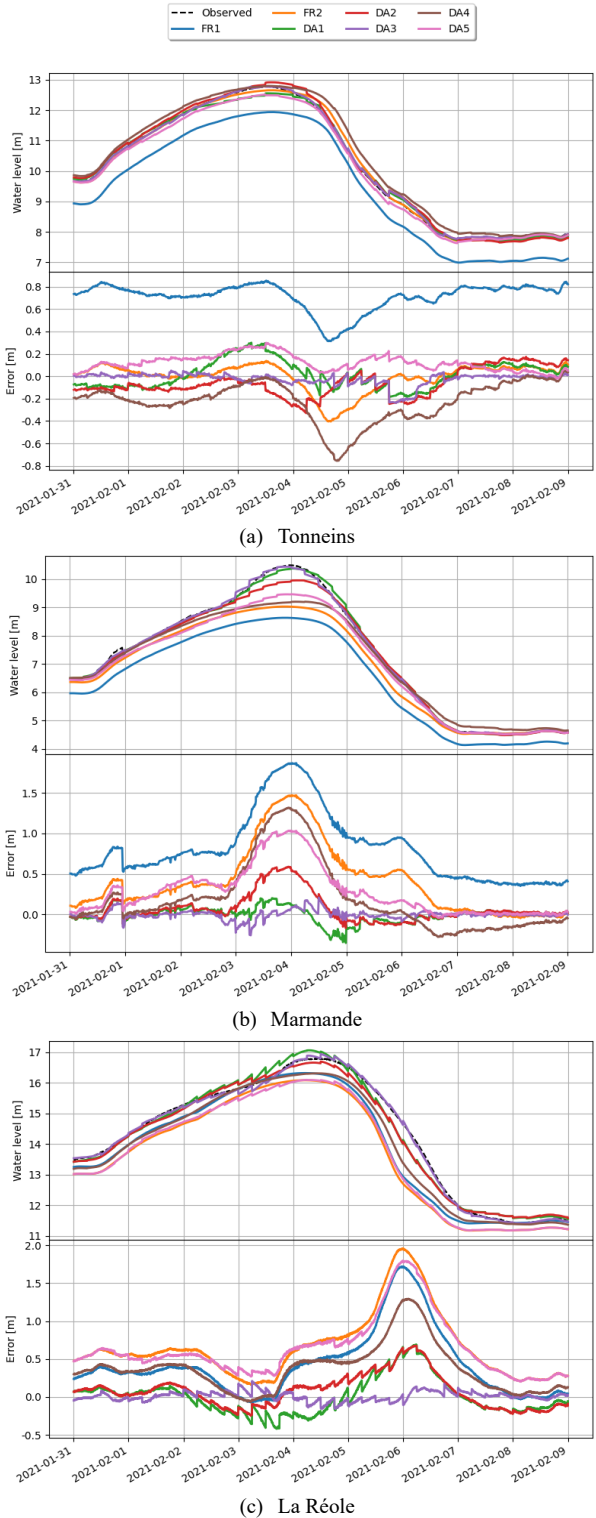

Figure 5: Simulated and observed time-series water levels around the flood peak, between January 31 and February 9.

## B. 1D Assessment Metrics

Table 3 summarizes the 1D assessment metrics (RMSE and maximum absolute error) computed over the whole simulation duration, from January 16 to February 15, between the water levels simulated by TELEMAC-2D and the observed water levels. For each metrics (each column) the best score (i.e., lowest RMSE and lowest maximum absolute error) is highlighted in boldface, whereas the second best is underlined.

Considering the free run experiments (FR1 and FR2), Table 3 shows that the bias correction leads to a better agreement between the model and observations at Tonneins and Marmande. This is however not the case for La Réole. Next, with the involvement of DA, the resulting water level errors have been reduced significantly. For example, from FR1 to DA1, the RMSE are reduced by 88.1%, 90.6% and 63.8% at Tonneins, Marmande and La Réole, respectively. Overall, DA3 yields the water levels the closest to the observations at every Vigicrue stations (as highlighted by the boldfaced values), both the RMSE and maximum absolute error, as a result of the $\sigma_{obs}$ set equal to 0.01 in which the observation vector was considered with very little perturbation (Eq. (9)). On the other hand, high uncertainty of observation vector is assumed for the DA4 experiment ($\sigma_{obs} = 0.99$), which leads to the resulting errors greater than DA2 and DA3.

In practice, an inflow discharge used as input for a hydraulic model like TELEMAC-2D may originate from a hydrologic model in which statistical corrections were also carried out [17]. This leads to the assumption that no uncertainty exists in the inflow discharge, based on which the DA5 was realized. However, the comparison between DA2 (with 7 parameters in the control vector) and DA5 (with only 4 friction coefficients in the control vector) demonstrates the improvements from DA5 to DA2 emphasized by the water level errors, which are reduced both in terms of RMSE and maximum absolute errors. This advocates for the consideration of uncertainty within $Q_{up}(t)$.

Table 3: 1D assessment metrics w.r.t. in-situ data measured at Vigicrue observing stations, computed over the whole simulation duration.

| Exp. name | Root-Mean-Square Error (m) | | | Max Absolute Error (m) | | |
|---|---|---|---|---|---|---|
| | Tonneins | Marmande | La Réole | Tonneins | Marmande | La Réole |
| FR1 | 0.756 | 0.625 | 0.409 | 1.062 | 1.870 | 1.721 |
| FR2 | 0.102 | 0.338 | 0.505 | 0.404 | 1.472 | 1.956 |
| DA1 | 0.090 | <u>0.059</u> | 0.148 | 0.456 | <u>0.352</u> | 0.690 |
| DA2 | <u>0.084</u> | 0.104 | <u>0.138</u> | 0.330 | 0.590 | <u>0.676</u> |
| DA3 | **0.034** | **0.034** | **0.043** | **0.238** | **0.258** | **0.212** |
| DA4 | 0.171 | 0.260 | 0.298 | 0.756 | 1.319 | 1.295 |
| DA5 | 0.093 | 0.226 | 0.480 | <u>0.299</u> | 1.036 | 1.796 |

## C. Control vector analysis

The analyzed values from DA experiments for friction and inflow parameters are shown in the first seven panels on Figure 6. The inflow discharge $Q_{up}(t)$ is also depicted on the eighth panel, and the two bottom panels reveal the water level errors from FR1 and FR2 at the Vigicrue stations. Since the model-observation bias is not taken into account in DA1, a larger increment is required in the control vector to make up for the difference between the model and observation. In order to make the model as close to the observation as possible, DA3 (with $\sigma_{obs} = 0.01$) commits very large correction upon the control vector, as shown by the large margin from the violet curves in all 7 control variable panels. Due to this constraint, the friction coefficient for the floodplain $K_{S_0}$ is even diminished below 10 $m^{1/3}s^{-1}$ near the flood peak which is not a realistic value for the floodplain in this catchment. On the other hand, DA4 with a $\sigma_{obs}$ set equal to 0.99, i.e., full uncertainty on the observation vector, presents very small corrections upon the control vector. As such, the brown curves remain quite stable from the calibrated and default values. Even though DA2 and DA5 experiments were configured with the same $\sigma_{obs}$, the respective increments on friction coefficients of DA5 (magenta curves) are much higher than those of DA2 (crimson curves). This is due to the fact that the control vector in DA5 is only composed of the friction coefficients, whereas DA2 involves all 7 variables, hence the contribution to compensate the water level error was shared among them.

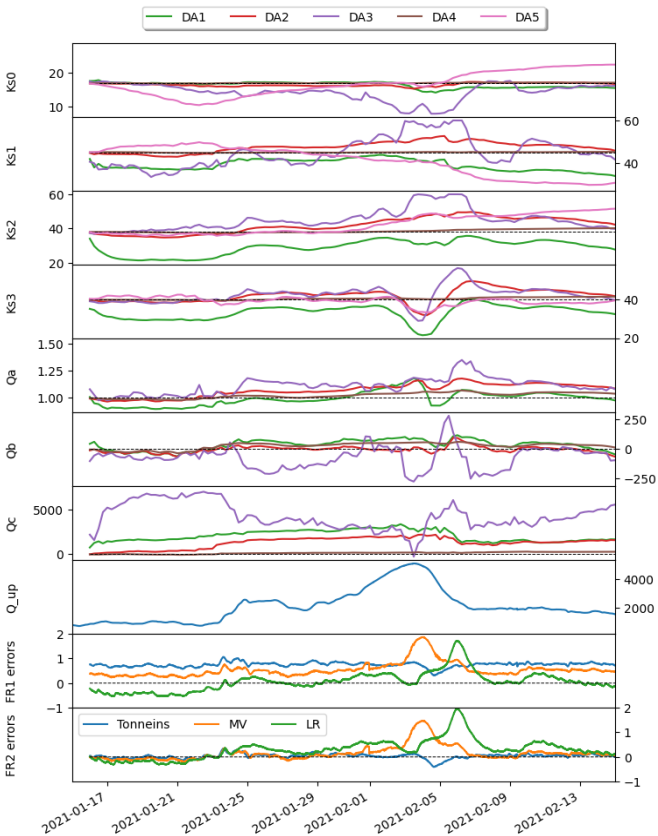

Figure 6: Evolution of controlled variables.

### D. Assessment of simulated flood extents

The flood extent maps simulated by TELEMAC-2D from the experiments are generated by applying a threshold of 5cm above which the node is considered wet (and dry otherwise). A rasterization of TELEMAC-2D water level output field onto a regular grid is also carried out. They are then compared with the flood extent maps generated from the RF algorithm on S1 images. Critical Success Index (CSI) is widely used to assess the performance of flood extent mapping, it is defined as follows:

$$\mathrm{CSI} = \frac{TP}{TP + FP + FN} \quad (13)$$

where True Positives ($TP$) is the number of pixels correctly predicted as flooded, False Positives ($FP$) or *over-prediction* is the number of non-flooded pixels incorrectly predicted as flooded, True Negatives ($TN$) is the number of pixels correctly identified as non-flooded, and False Negatives ($FN$) or *under-prediction* is the number of missed flooded pixels. The RF-inferred flood extent maps are considered the reference flood map based on which the TELEMAC-2D flood extent map will be evaluated.

Table 4 summarizes the CSI measured from the experiments with respect to the RF-inferred flood extent maps. On each day, the best score (i.e., highest CSI) is highlighted in boldface, whereas the second best is underlined. At the beginning of the event, the water prevails mainly in the river bed and only occupies a small portion of the simulation domain. However, the CSI scores during these non-flooding times were degraded because of several over-predicted regions such as the numerical artificial flooding of the upstream first meander (previously mentioned in subsection II.B). Let us focus on the 2021-02-03 which is the S1 overpass time the nearest to the flood peak. DA4 with the most uncertainty hypothesized on the observation vector allows the highest CSI score. However, reducing the $\sigma_{obs}$ from 0.99 (DA4) to 0.15 (DA2) only decreases the resulting CSI 0.13 points (i.e., from 63.97% to 63.84%). DA3, by forcing the simulated water levels to be as close to the observation as possible, yields significantly lower CSI, even smaller than the free runs. After the flood peak, the interpretability of the CSI may also be limited due to a number of water puddles that remain in the floodplain as the model struggles to simulate the water recession.

Table 4: CSI metric with respect to S1-derived flood extent maps.

| | CSI (%) | | | | | |
|---|---|---|---|---|---|---|
| Exp. name | Jan 26 07:00 | Jan 27 19:00 | Jan 28 19:00 | Feb 2 19:00 | Feb 3 19:00 | Feb 7 7:00 |
| FR1 | **28.13** | **27.81** | **25.09** | 44.39 | 55.86 | 22.10 |
| FR2 | | | | | | |
| DA1 | 25.82 | 25.92 | 23.58 | 41.91 | 61.97 | 20.00 |
| DA2 | 27.66 | 27.49 | 24.95 | 43.98 | 63.84 | 20.87 |
| DA3 | 27.65 | 27.49 | 24.93 | 44.22 | 50.82 | 23.75 |
| DA4 | 27.52 | 27.44 | 24.84 | **47.08** | **63.97** | 20.92 |
| DA5 | 27.74 | 27.59 | 24.92 | 44.60 | 49.31 | **25.41** |

### E. Analyzing local behavior at virtual observation regions

Several locations within the simulation domain were selected to analyze local behavior of the flood simulation, especially in the floodplains, as shown by the rectangular boxes in Figure 7. From these boxes, the number of wet pixels is counted from the flood extent maps (there are 9 in total, cf. Figure 2). These counts from the experiments (represented by the dashed curves) and the RF-inferred flood extent maps (black solid curve) are depicted in Figure 8.

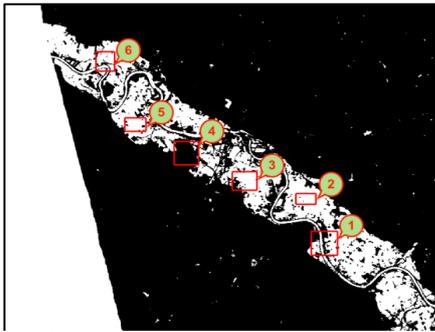

Figure 7: Virtual observation regions overlapped on a RF-inferred flood extent map, generated from S1 image on February 3.

All of virtual observation regions are filled very quickly with respect to the S1 observations. One particular region is box number 4 in which the number of flooded pixels is very low according to the observation, but it is over-flooded in the experiments, with varying degree (FR2 and DA4 by a small margin, and the others by a lot). Such different behaviors between boxes number 3, 4, and 5, despite they are all on the floodplain and near to each other, advocate for an adjustment, e.g., DA, which corrects the friction of the area around box number 4. DA1, without taking into account the bias correction, tends to overflood the whole catchment (i.e., for all boxes). Lastly, the flood recession period associated with the last two S1 overpass times is poorly modelled, particularly for box number 2. This stems from the fact that evaporation and ground infiltration physical processes are not accounted for in the Garonne model.

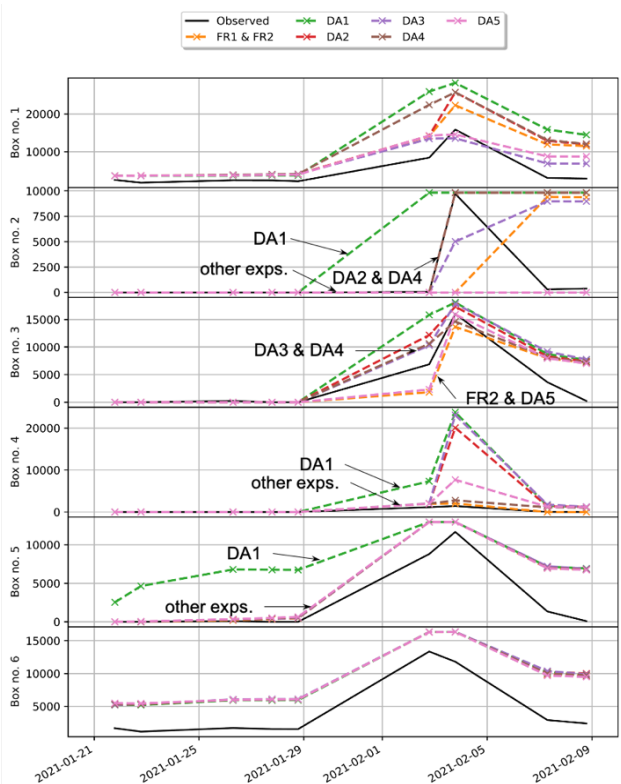

Figure 8: Wet pixel counts at virtual observation regions.

## V. Conclusions

In this paper, flood extents observed from Sentinel-1 images were extracted using Random Forest and compared with the flood extent maps simulated by TELEMAC-2D. The study was carried out over the Garonne Marmandaise catchment, focusing on the flood event occurred in January and February 2021. Seven experiments were realized, namely two in free run mode and five in DA mode. The DA was performed using only in-situ observations, and it was implemented by an EnKF with a 12-hour assimilation window sliding with 6-hour overlapping between windows.

Several key remarks can be drawn from this work. First, ensemble-based data assimilation allows time-varying correction of friction and inflow leading to improved simulation and forecast in the river bed and the flood plain. Second, it was shown that the bias correction leads to properly corrected water levels in the river bed and floodplains. The uncertainty assumed on the observation vector through parametrization of $\sigma_{obs}$ was also demonstrated, for which the compromise between a good fit on simulated water level (i.e., DA3) and a high CSI/good agreement between flood extents (i.e., DA4). Lastly, several limitations concerning local behaviors of the flood simulation have been revealed with the virtual observation regions.

Moving forward, a particular perspective for this study concerns resolving the limitation of the control vector size and refining the spatial friction zoning definition, assimilating information from RS data such as number of wet pixels in the flood plain (which were used for validation in this work) and/or flood extent information to improve the description of the refined friction subdomains in the floodplains and to calibrate them. This study paves the way toward a cost-effective and reliable solution for flood forecasting and flood risk assessment over poorly gauged or ungauged catchments, thanks to the use of remote sensing data. Such developments, once generalized, could potentially lead to hydrology-related disaster risk mitigation in other regions. Future progresses built upon this work will involve a more refined approach for friction zoning and calibration, especially for the floodplains, as well as improving the current models by assimilating flood extent maps.

## Acknowledgement

This work was supported by CNES, CERFACS and SCO-France. The authors gratefully thank the Electricité de France R&D for providing the TELEMAC-2D model for the Garonne River, SCHAPI, SPC Garonne-Tarn-Lot and SPC Gironde-Adour-Dordogne for providing the in-situ data. Lastly, the authors would like to thank the anonymous reviewer whose comments and suggestions helped improve this manuscript.